\documentclass[10pt,aps,preprint]{revtex4}
\usepackage{float}
\usepackage{graphicx}
\usepackage{color}
\usepackage{multirow}
\begin{document}
\title{Band gap modulation in polythiophene and polypyrrole-based systems}
\author{Thaneshwor P. Kaloni$^1$}
\email{thaneshwor.kaloni@umanitoba.ca}
\author{Georg Schreckenbach$*^1$}
\email{schrecke@cc.umanitoba.ca}
\author{Michael S. Freund$^2$}
\email{msfreund@fit.edu}
\affiliation{$^1$Department of Chemistry, University of Manitoba, Winnipeg, Manitoba R3T 2N2, Canada}
\affiliation{$^2$Department of Chemistry, Florida Institute of Technology, 150 West University Boulevard, 
Melbourne, Florida 32901-6975, USA}
\begin{abstract}
In this paper, the structural and electronic properties of polythiophene and polyprrrole-based systems have been investigated using first-principles calculations both in periodic and oligomer forms. Of particular interest is the band gap modulation through substitutions and bilayer formation. Specifically, S has been substituted by Se and Te in polythiophene, leading to polyseleophene and polytellurophene, respectively, and N has been substituted by P and As in polypyrrole. The values obtained of the binding energy suggest that all the systems studied can be realized experimentally. Stacking (bilayer formation) of pure polythiophene, polypyrrole and their derivatives leads to linear suppression of the band gap or HOMO-LUMO gap as a function of the stacking. Mixed bilayers, including one formed from polythiophene on top of polypyrrole, have also been considered. Overall, a wide range of band gaps can be achieved through substitutions and stacking. Hybrid (B3LYP) calculations also suggest the same trend in the band gap as PBE calculations. Trends in the binding energy are similar for both periodic and molecular calculations. In addition, the $\Gamma$-point phonon calculation are performed in order to check the stability of selected systems.
\end{abstract}
%\pacs{73.22.Gk, 75.50.Pp, 73.43.Cd}
%\keywords{Polythiophene, Polypyrrole, Electronic properties, Binding energy, Substitution doping}
%\begin{document}
\maketitle
\section*{Introduction}
In 1976, $\pi$-conducting polymers were discovered by Heeger, MacDiarmid, Shirakawa and co-workers. This discovery opened the way for comprehensive investigations and understanding of various aspects of the $\pi$-conducting polymers from both physical and chemical points of views.\cite{1} The $\pi$-conjugated polymers possess alternating single and double bonds of carbon atoms leading to one $\pi$ electron for each of the carbon atoms.\cite{2,3,4,5,6} The carbon atoms are $sp^2p_z$ hybridized and overlap along the polymer chain, inducing electron delocalization in the polymer chain.\cite{7,8} The electronic delocalization opens up a path for charge transport along the polymer chain, and as a consequence, the polymer chain can have conducting or semiconducting nature.\cite{9,10,11,12,13,14,15,16,17,18,19,20,21,22,23} 
Organic $\pi$-conjugated polymers have been investigated widely in experiment and theory. These polymers are of great interest because of their potential to be integrated into electronic and optical devices \cite{1,24,25,26,27,28,29,30,31,32,33,34,35,36,37,38,39,40} as well as sensors.\cite{4,41,42,43,44,45,46}

It has already been reported that the electronic properties of such polymers can be engineered/tuned without much efforts either by chemical modification or atomic/molecular scale doping. \cite{8,46,47,48,49} Polyacetylene can be doped at room temperature by various dopants over a wide range of concentrations such that the structural and electronic properties can be tuned.\cite{1,50} Doping has been widely utilized for various conjugated polymers. It was found that dopants play a crucial role in changing their properties compared to those of conventional semiconductors.\cite{46} The doping can be achieved either by charge transfer or by application of an external electric field. By applying these doping 
techniques, the electronic and optical properties of conducting polymers can be engineered widely, including the transition from semiconductor to metal or further to insulator depending upon the dopant concentration.\cite{48,51,52,53} This induces an outstanding opportunity for switching between conducting and insulating properties, which indeed paves the way for applications in optoelectronics, such as organic polymer based transistors, photoresistances, light-emitting diodes, and polythiophene based organic solar cells.\cite{54}

There have been several efforts to address the electronic structure of polythiophene derivatives using first-principles calculations \cite{53,55,56,57,58,59,60,61,62} Substitution in polythiophenes, such as replacement of H by CH$_3$, NH$_2$, NO$_2$, and Cl, has been studied. It was found that the electronic structure can easily be tuned.\cite{62} The optical properties of fully conjugated cyclo[n]thiophenes have been studied in the experiment and theory and it has been demonstrated that the band gap and optical properties depend on the structure of the thiophene molecules.\cite{63,64} In addition, thiophene and its derivatives are expected to be utilized in electronic memory devices. These materials have been in focus for the last two decades.\cite{54,62} Injection of electrons from the highest occupied molecular orbital (HOMO) to the lowest unoccupied molecular orbital (LUMO) is achieved without difficulties in this class of materials.\cite{65} Recently, synthesis of polyselenophene/polytellurophene (replacement/substitution of the S atoms by Se/Te) has been achieved.\cite{66,67,68,69,70} The material can be used as an active layer in field-effect transistors.\cite{69} Moreover, in a recent experiment, it has been reported that such systems are promising for CO$_2$ capture.\cite{71}

Using a conjugated polymer, a well-established barrier to Li-ion drift has been achieved experimentally using polythiophene and polypyrrole, which in fact provides a mechanism for creating high-performance memory devices.\cite{72} For such a system, the conjugated polymer barrier layer made of polypyrrole remains conducting at the interface between the metal oxide and Li$^+$ ion doped polythiophene. Moreover, the structural and electronic properties of polythiophene doped by Li/Cl in various stacking schemes in both periodic and oligomer forms have been investigated.\cite{53} It has been suggested that these types of study will be useful for understanding the structural properties and the tunability of the electronic states, which should be an important step toward construction of polythiophene-based electronic devices. In this paper, we investigate the structural and electronic properties of S substituted by Se and Te in polythiophene and N substituted by P and As in polypyrrole using first-principle density functional theory (DFT) based calculations. It is believed that the comprehensive analysis of our investigation will be very useful for understanding the structural properties and the band gap or HOMO-LUMO gap of the systems under study, which provides a basis for constructing polythiophene and polypyrrole-based electronic devices. Moreover, in case of the periodic calculations, we also performed hybrid (B3LYP) calculations in order to determine the effect of the hybrid functional on the band gap. Furthermore, the $\Gamma$-point phonon calculations were performed to confirm the stability of the selected systems.

\section*{Methods}
All the periodic calculations were performed using DFT within the generalized gradient approximation in the Perdew, Burke, and Ernzerhof (PBE) parametrization \cite{73} as implemented in the Quantum-ESPRESSO code.\cite{74} The effects of spin-orbit coupling (spin-orbit relativistic effects with PBE pseudopotential) were also taken into account. For the systems with stacking configurations, the van der Waals interactions (DFT-D) were included in order to obtain correct dispersion.\cite{75} A relatively high plane wave cutoff energy of 800 eV with a larger Monkhorst-Pack $32\times1\times1$ k-mesh were employed. The considered supercell contains four monomers of thiophene as well as pyrrole and has lattice constants of $a=15.59$ \AA\ and $b=15.00$ \AA. A vacuum layer of 20 \AA\ was used in order to nullify any spurious interactions due to the periodic boundary conductions. The atomic positions were optimized very well with all the forces converged to 0.0005 eV/\AA. The following systems were considered with a vacuum along the $y$- and $z$-directions (see Tables 1-4) (i) monolayer/bilayer of pure polythiophene, (ii) monolayer/bilayer of polythiophene substituted by Se (polyselenophene), (iii) monolayer/bilayer of polythiophene substituted by Te (polytellurophene), (iv) monolayer/bilayer of pure polypyrrole, (vi) monolayer/bilayer of polypyrrole substituted by P, (vi) monolayer/bilayer of polypyrrole substituted by As, and (vii) bilayer made of polythiophene and polypyrrole. In addition, various mixed bilayers were considered as summarized in Tables 1-4. All of the above systems were also considered for the molecular case. In addition, the B3LYP hybrid functional \cite{76} was employed in order to examine the effect of the hybrid functional on the band gap. Note that the B3LYP calculations were performed only for the periodic cases and not extended to the molecular cases due to limitations that in computational resources. We believe that the trends in the band gap would be that the same in molecular and periodic calculations. 

In the molecular calculations (hexamers), the systems were optimized using DFT-D (PBE) \cite{73} as implemented in the Amsterdam Density Functional (ADF) code.\cite{77,78,79} Previously, we have shown that this setup provides reasonable results while comparing to the corresponding experimental findings.\cite{53} Note that experimentally and theoretically, oligomers of six or less thiophene rings are found to be sufficient to reasonably model such systems.\cite{80} All the calculations were performed using double zeta Slater-type orbital basis sets (ZORA-DZ). Relativistic effects including spin-orbit effects were included (ZORA with spin-orbit) \cite{81,82,83}, they cannot be neglected due to larger values of the spin-orbit coupling and scalar relativistic effects especially for Se and Te atoms. All the geometries were optimized until the energy was converged to 0.0003 eV. A default integration accuracy parameter of 4.0 was used for geometry optimizations and single point calculations for all systems under study.

\section*{Results and discussion}%polythiophne-phonon.ps
\subsection*{Polythiophene and its derivatives (Periodic approach)}

\subsubsection*{Vibrational properties}
\begin{figure*}[htbp]
\includegraphics[width=10cm]{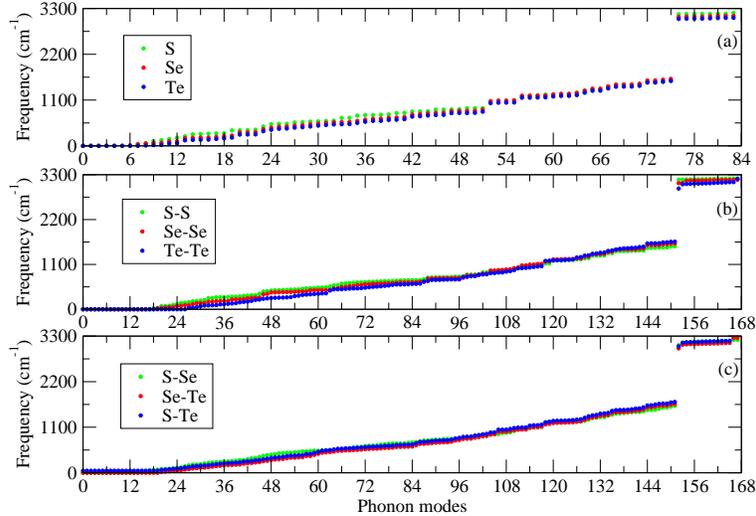}
\caption{$\Gamma$-point phonon modes of polythiophene derivatives (a) monolayers (S, Se, and Te), (b) identical bilayers (S-S, Se-Se, Te-Te), and (c) non-identical bilayers (S-Se, S-Te, and Se-Te), see the optimized structures depicted in Fig.\ 3(a) and left panel of Fig.\ 4.}
\end{figure*}
Vibrational Raman and infrared frequencies are really useful to provide information on a molecular or atomic level. The vibrational frequencies can easily be detected in experiment and the frequencies can vary between different materials as well as due to distortions in the structures arising from doping or defects.\cite{84,85,86} Therefore, first of all, we performed series of phonon calculation at the $\Gamma$-point phonon for polythiophene and its derivatives in order to check the structural stability. Indeed, only stable structures are presented in this paper as listed in Table I. Each atom has three degrees of vibrational freedom, each monolayer of polythiophene and its derivatives have a total of 28 atoms in the system, thus, the total number of phonon modes in these structures is 84, see Fig.\ 1(a), whereas the number of phonon modes becomes 168 for identical and non-identical bilayers (28 atoms in each layer), see Figs.\ 1(b-c). The highest phonon frequency for a monolayer of polythiophene is found to be 3195 cm$^{-1}$ at the $\Gamma$-point, which agrees well with an experimental observation.\cite{84} Note that no negative phonon frequencies are observed at the $\Gamma$-point for all the system under study, which confirms that the systems are stable. The highest phonon frequency for monolayers of polyselenophene and polytellurophene is 3124 cm$^{-1}$ and 3077 cm$^{-1}$ at the $\Gamma$-point, respectively. Our obtained values show good agreement with a previous theoretical prediction.\cite{87} It is clear that the phonon frequencies are decreasing while moving from polythiophene to polytellurophene (S to Te), which is called phonon softening \cite{88}. The phonon softening occurs due to the fact that the weakening of C-Se and C-Te bond lengths as compared to C-S bond lengths. However, the phonon frequencies become harder in case of identical as well as non-identical bilayers as compared to their monolayer counterparts because the interlayer interactions (out-of-plane vibrations) play a pivotal role. The obtained value of the phonon frequencies for the bilayers are found to be 3207 cm$^{-1}$, 3178 cm$^{-1}$, 3198 cm$^{-1}$, 3210 cm$^{-1}$, 3260 cm$^{-1}$, and 3310 cm$^{-1}$ for S-S, Se-Se, Te-Te, S-Se, Se-Te, and S-Te (see Figs.\ 1(b-c), respectively.

\begin{figure*}[t]
\includegraphics[width=10cm]{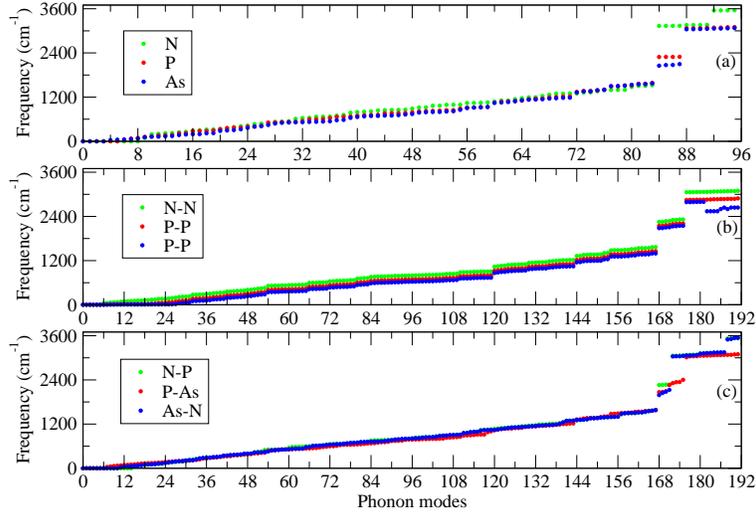}
\caption{$\Gamma$-point phonon modes of polypyrrole derivatives (a) monolayers (N, P, and As), (b) identical bilayers (N-N, P-P, and As-As), and (c) non-identical bilayers (N-P, N-As, and P-As, see the optimized structures presented in the Fig.\ 3(b) and right panel of the Fig.\ 4.}
\end{figure*}
Furthermore, similar to the polythiophene derivatives, we also performed a series of phonon calculation at the $\Gamma$-point for polypyrrole and its derivatives to check the structural stability. Only stable structures are presented in Table I, Fig.\ 1(b). Each monolayer of polypyrrole and its derivatives has a total of 32 atoms in a unit cell; therefore, the total number of phonon modes in these structures is 96, see Fig. 2(a). But the number of phonon modes becomes 192 for identical and non-identical bilayers (32 atoms in each layer) as depicted in the Figs.\ 2(b-c). We observe a highest phonon frequency of 3559 cm$^{-1}$ at the $\Gamma$-point for the monolayer of polypyrrole, which agrees well with a previously reported value.\cite{89} No negative phonon frequencies are observed at the $\Gamma$-point for all the system under consideration, which confirms that the structures presented in the paper are stable. The highest phonon frequency for P and N amounts to 3103 cm$^{-1}$ and 3074 cm$^{-1}$ at the $\Gamma$-point. It should be noted that we do not have any previously published data to be compared with our results for these systems. Similarly to the polythiophene, it is clear that the phonon frequencies are decreasing while moving from N to P to As; the weaker bonding is responsible for the softening of the phonon as we move from N to P to As, see Fig.\ 2(a). However, the phonon frequency becomes softer/harder in case of identical/non-identical bilayers as compared to their monolayer counterparts because the interlayer interactions (out-of-plane vibrations) play a crucial role. In contrast to polythiophene bilayers, the phonon frequencies in case of identical bilayers become softer due to the fact of a strong deformation in the atomic structure in both of the layers, while non-identical bilayers follow the same trend as the polythiophene derivatives(non-identical bilayers). The calculated values of the highest phonon frequencies for bilayers are found to be 3094 cm$^{-1}$, 2828 cm$^{-1}$, 2640 cm$^{-1}$, 3533 cm$^{-1}$, 3195 cm$^{-1}$, and 3541 cm$^{-1}$ for N-N, P-P, As-As, N-P, P-As, and As-N (see Figs.\ 2(b-c)), respectively.

\subsubsection*{Structural arrangements}

\begin{figure*}[htbp]
\includegraphics[height=9cm]{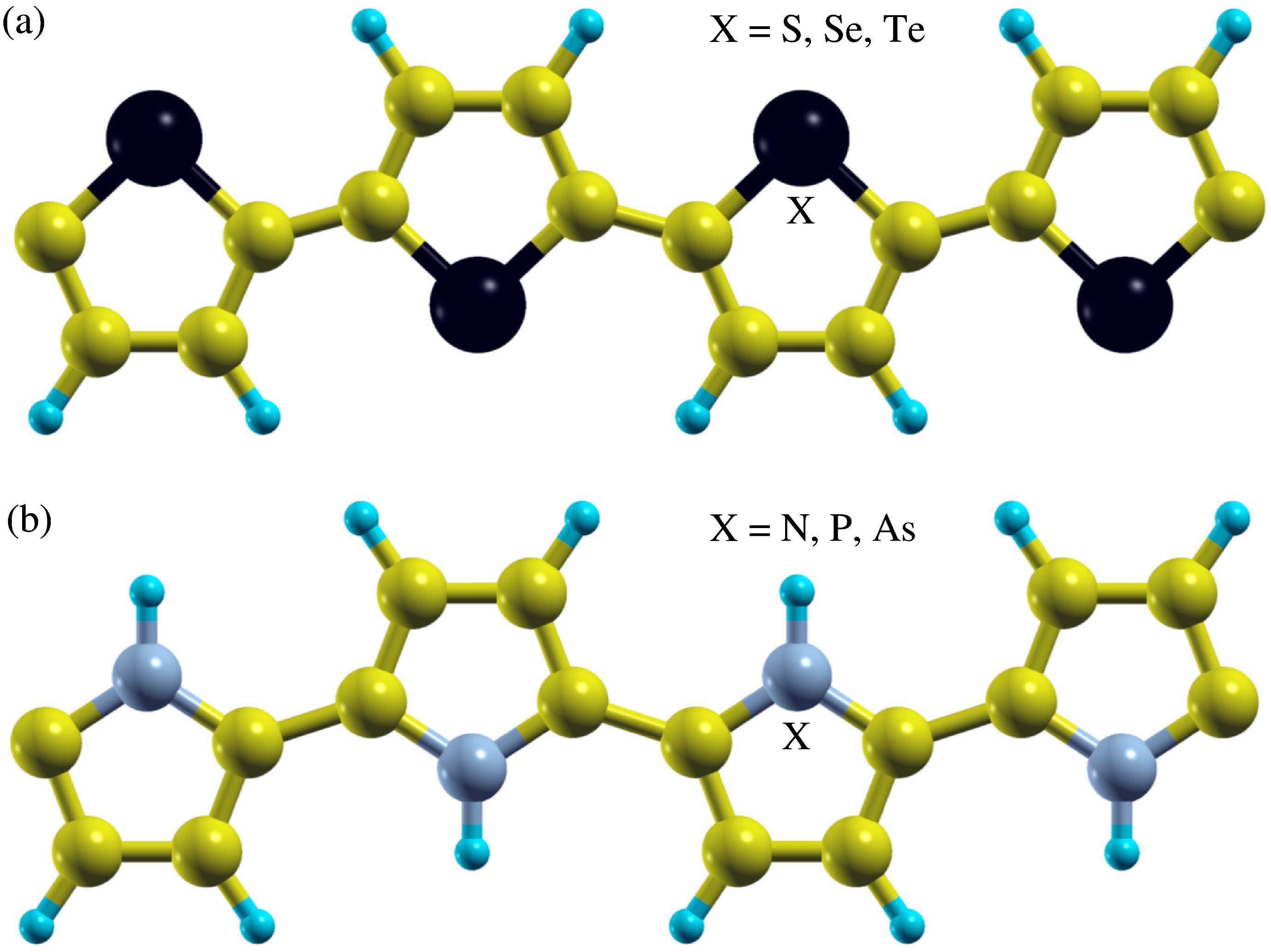}
\caption{Top view of the structures (periodic approach) of (a) polythiophene  and (b) polypyrrole based materials, where yellow, blue, black, and gray balls represent C, H, S or Se or Te, and N or P or As atoms, respectively.}
\end{figure*}

The structure of monolayer polythiophene is shown in Fig.\ 3(a). The polyselenophene and polytellurophene have very similar structures. Side views of the optimized structures of various bilayers are presented in Fig.\ 4. The considered supercell of the monolayer has 16 carbon (C) atoms, 8 Hydrogen (H) atoms, and 4 Sulphur (S) or Selenium (Se) or Tellurium (Te) atoms. The structural parameters, which consist of various bond lengths and bond angles, are summarized in Table 1. The values obtained of bond lengths and bond angles agree well with the available experimentally and theoretically obtained values.\cite{53,66,90} Polythiophene, containing S atoms in the pentagonal rings of the polymer backbone, has been well studied in both experiment and theory, whereas the effect of Se or Te instead of S is not well understood. Recently, replacement/substitution of S by Se has been successfully realized experimentally, resulting in polyselenophene.\cite{53,67,91} It has been found that such a substitution is crucial in order to control the electronic properties of the polythiophene based polymer, especially the band gap. 

For pristine polythiophene, the buckling, which is defined as the perpendicular out-of-plane distance between atoms in the same layer, is found to be 0.10 \AA\ (Table 2), C$-$C bond lengths are found to be 1.38-1.41 \AA, and the C$-$S bond length amounts to 1.73 \AA. While the C-S-C, C-C-C, and S-C-C bond angles are found to be 93$^\circ$, 113-128$^\circ$, and 109-119$^\circ$, respectively. The calculated structural parameters for polyselenophene are found to be similar to those of polythiophene except for the C$-$Se bond length, which grows as compared to the C$-$S bond length in polythiophene and is found to be 1.88 \AA. At the same time, the C$-$C bond length shrinks slightly, in good agreement with the experimentally obtained value.\cite{61}

Our calculated values for the structural parameters of polytellurophene are listed in Table 1 as well. The buckling is found to be 0.17 \AA, slightly increased as compared to polythiophene (0.10 \AA) and polyselenophene (0.12 \AA). The C$-$C bond lengths shrink to 1.35-1.38 \AA\ as compared to the C$-$C bond lengths in polythiophene and polyselenophene, while the C$-$Te bond length grows to be 2.08 \AA. All other parameters are similar to those of the previous cases. It should be noted that C$-$C bond lengths decrease from S to Te and the C$-$X (X = S, Se, Te) bond length grows from Se to Te. Note that the exact structural modification can be seen from moleculer dynamics study, which indeed is close to the experimental observation as recently proposed by Salvador and co-workers.\cite{salvador} %Interestingly, experimentally it is suggested that replacement/substitution can be applied to other systems, such as Si or Ge in place of C, P in place of N as Se or Te in place of S. However, such an experiment may require huge synthetic efforts and carefulness.\cite{C5TC00158G} This type of element-based conjugated polymers are expected to be promising for application in optoelectronic devices.\cite{C3RA40286J} 

In addition, we have studied bilayers of polythiophene, polyselenophene, and polytellurophene in two scheme (i) both layers are identical (S-S, Se-Se, and Te-Te) and (ii) the two layers are not identical (S-Se, S-Te, and Se-Te). The side views of the optimized structures are depicted in Fig.\ 4 (left side). The interlayer spacings for scheme (i) are found to be 3.33-3.43 \AA, 3.36-3.44 \AA, and 3.36-3.42 \AA, for polythiophene, polyselenophene, and polytellurophene, respectively. %The interlayer spacing agrees well with previous results in case of pristine polythiophene bilayers.\cite{kaloni-jpcc2015} 
The buckling is found to increase from polythiophene to polytellurophene and the C$-$C bond lengths are found to decrease from polythiophene to polytellurophene, respectively. Whereas the C$-$X bond length is found to be increasing from polythiophene to polytellurophene, almost identical to the monolayers. The other structural parameters vary in a similar pattern as in case of the corresponding monolayers, compare Table 1. 

\begin{figure*}[htbp]
\includegraphics[height=9cm]{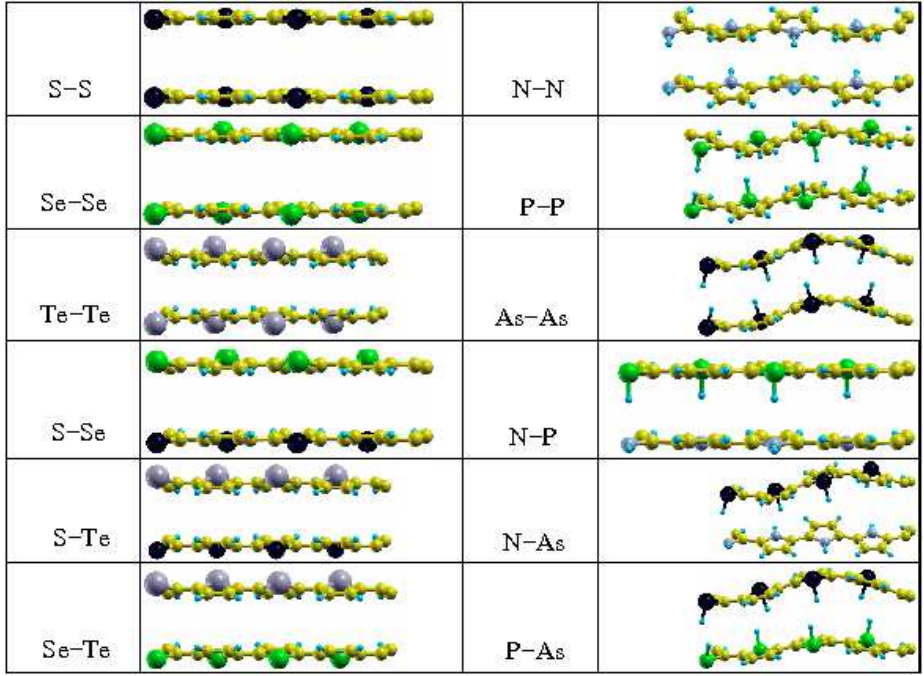}
\caption{Optimized bilayer structures, periodic calculations. Yellow, blue, black, green, and gray balls represent C, H, S/As, Se/P, and Te/N atoms, respectively.}
\end{figure*}

The interlayer spacing amounts to 3.42-3.62 \AA, 3.39-3.65 \AA, and 3.39-3.68 \AA, respectively, in case of scheme (ii). Importantly, the buckling is quite large as compared to that of scheme (i); the probable reason for the increment in the buckling is the interaction between the X atoms, e.g. the S atom of one layer with the Se atom of another overlying layer. This is not the case for scheme (i) because both the layers are identical. In addition, C$-$S, C$-$Se, and C$-$Te bond lengths of 1.73-1.88 \AA, 1.74-2.06 \AA, and 1.88-2.07 \AA\ are found for S-Se, S-Te, and Se-Te bilayers, respectively, again this is different from scheme (i). 

Furthermore, polymers that contain alternating S, Se and Te atoms in consecutive rings have also been considered, either in a monolayer or bilayer form (referred to S-Se-Te$^1$ and S-Se-Te$^2$ in Tables 1-4). It is found that the buckling in the bilayer is about 6 times larger than that of the monolayer because of multi-atomic interactions, i.e. S-S, Se-Se, and Te-Te atomic level interactions (note that AA stacking has been considered). The remaining structural parameters are consistent with the other systems discussed above. Based on our calculations, such systems are promising to reduce the band gap, which will be discussed in detail below.     

\subsubsection*{Electronic structure}
\begin{figure*}[htbp]
\includegraphics[width=0.9\textwidth,clip]{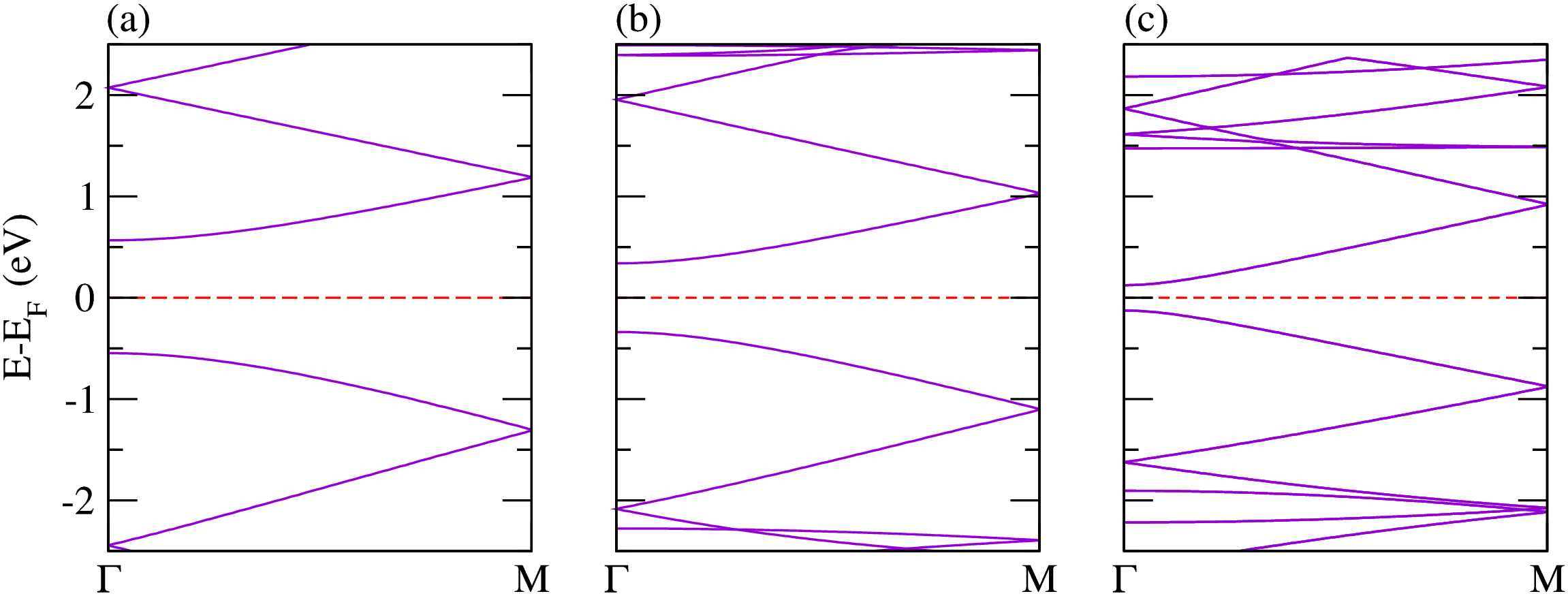}
\caption{The electronic structure of single layers of (a) polythiophene, (b) polyselenophene, and 
(c) polytellurophene.}
\end{figure*}

Polythiophene and its derivatives are a class of materials that have been widely used as conjugated materials in electronic devices such as optoelectronics.\cite{91,92} Nevertheless, there are certain difficulties to integrate polythiophene in devices mainly due the limited absorption profile. In order to improve device performance, lots of research activities have been focused on the development of new low band gap polymers made from polythiophene and its derivatives. In Fig.\ 5, the electronic structures of polythiophene, polyselenophene, and polytellurophene are addressed. The calculated value of the valence bandwidth (which is defined as the energy difference between the first band located right below the Fermi level at $\Gamma$-point to the M-point) is found to be almost constant at about 0.74 eV, whereas the conduction bandwidth (which is defined as the energy difference between the first band located right above the Fermi level at $\Gamma$-point to the M-point) increases significantly from 0.63 eV to 0.69 eV and 0.81 eV, respectively for polythiophene, polyselenophene, and polytellurophene, see Table 2. Concurrently, the value obtained of the band gap is reduced from 1.12 eV to 0.68 eV to 0.26 eV for polythiophene, polyselenophene, and polytellurophene, respectively, compare Fig.\ 5(a-c). Whereas the experimental value of the band gap for polythiophene was found to be 2.0 eV, which indeed agrees well with our calculated band gap in the framework of the hybrid (B3LYP) calculations of 2.08 eV. Using the same functional, the obtained values of the band gap are 1.53/0.97 eV for polyselenophene/polytellurophene. It has already been established that the band gaps obtained from the hybrid (B3LYP) calculations reproduce the experimental band gaps very well.\cite{61} Note that the trend in the band gap using hybrid (B3LYP) calculations (reduction in the band gap) from polythiophene to polytellurophene and their derivatives follows the same trend as the PBE calculations, see Table 2. The values obtained of the band gap for polythiophene and polyselenophene agree well with previous reports \cite{62,90,93}, however, there is no previous experimentally/theoretically obtained value of the band gap for polytellurophene to be compared. Note that replacing the S atom in polythiophene by heavier atoms such as Se or Te results in lowering the band gap. Previous studies indicated that the HOMO is not contributed by the S atom; on the other hand the LUMO has significant contributions from the S atom.\cite{91,94} Thus, replacement of S by Se or Te should have a large contribution to the LUMO. Because of the lower ionization potential of the Se or Te atoms, their incorporation into polythiophene should lead to a lowered LUMO energy level, and hence, provide for a smaller energy gap as compared to its S counterpart.

Importantly, lowering the band gap along with increasing the conduction bandwidth make the material a good candidate to be $n$-doped, and it should possess a large $n$-type conductivity as compared with the systems with larger band gap.\cite{95} Moreover, the conductivity, ionization potential, and electron affinity should be largest for polytellurophene amongst the three materials. Similarly, the effective masses of holes and electrons in the valence as well as conduction bands should follow the same trend as conductivity, ionization potential, and electron affinity. The calculated value of the binding energy [$E_{binding} = (E_{supercell}-E_X-E_Y-E_Z)/N$, where $E_{supercell}$, $E_X$, $E_Y$, and $E_Z$ are the total energy of the supercell, total energy of isolated C atoms, total energy of isolated S or Se or Te atoms, and total energy of isolated H atoms, while $N$ is the total number of atoms in the system] decreases slightly from polythiophene ($-0.483$ eV) to polytellurophene ($-0.455$ eV). Note that, essentially, all the three systems should be easily realized experimentally because of only slight differences in the binding energy. Likewise, the nature of the bonding is expected to be quite similar in the whole network/backbone. Indeed, polytellurophene and polyselenophene have been synthesized.\cite{34,38,39,69,96,97,98}

\begin{figure*}[htbp]
\includegraphics[width=0.9\textwidth,clip]{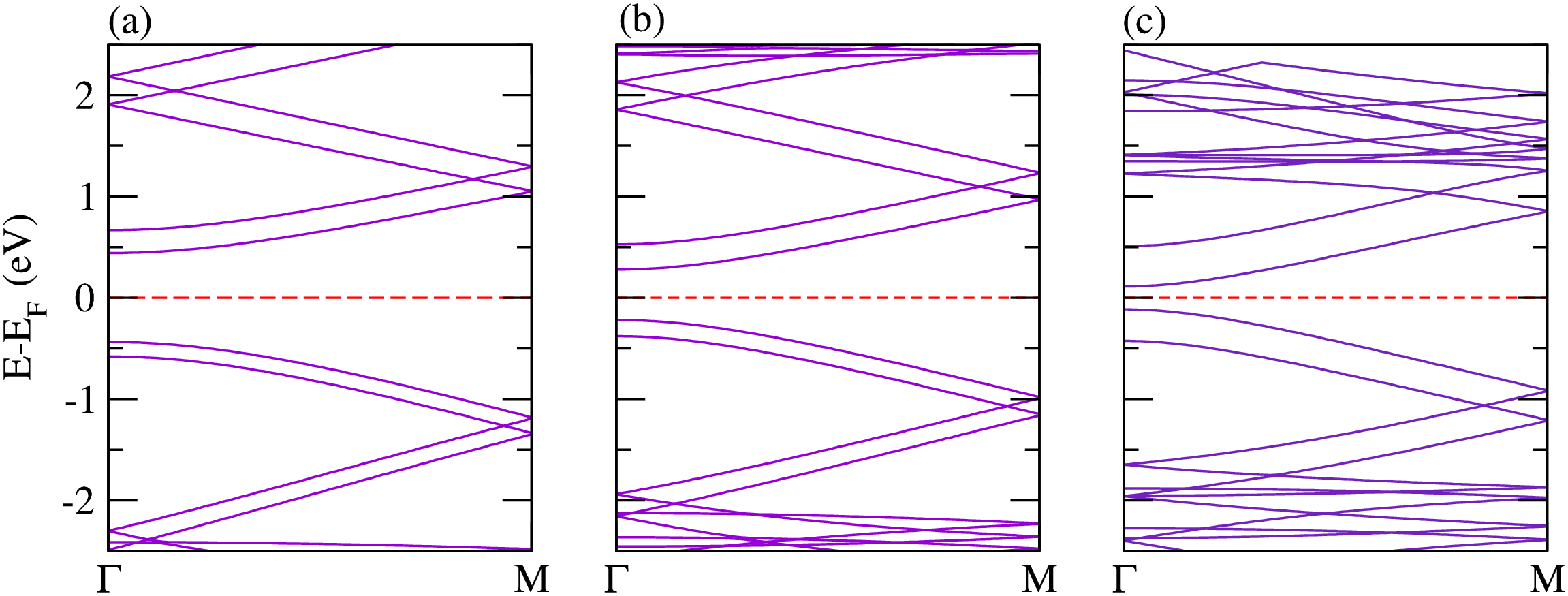}
\caption{The electronic structure of bilayers of (a) polythiophene, (b) polyselenophene, and (c) polytellurophene.}
\end{figure*}

The electronic band structures of bilayer polythiophene, polyselenophene, and polytellurophene, where both the layers are identical, are addressed in Fig.\ 6. The calculated value of the valence bandwidth decreases from polythiophene to polytellurophene, while the conduction bandwidth increases, similar to the monolayer counterparts, see Table 2. The band gap is reduced substantially as compared to the monolayer counterparts; the reduction in the band gap can be understood as a result of the interaction between the two identical layers\cite{53}. The obtained value of the band gap using the hybrid (B3LYP) functional becomes larger than the value obtained form the PBE functional, as expected, and, importantly, the band gap trend is similar as compared to the PBE functional, see Table 2. A number of synthetic routes have been proposed experimentally to produce various low band gap polymers using bulk or bilayer schemes, including donor-$\pi$-acceptor conjugated polymers, for example \cite{99}, as well as polythiophenes grafted with side chains \cite{100} copolymerized with fluorene moieties.\cite{101} Such low band gap materials are promising for harvesting solar photons.\cite{102} Therefore, our results point to new forms of conducting polymers that would be useful in solar cells. Moreover, a sizeable band gap is found for two immediate valence and conduction bands; the values of the band gap in the valence as well as conduction bands edges are found to be increasing from polythiophene to polytellurophene, see the data summarized in Table 2. Importantly, there are two possible optical transitions during light absorption and exciton formation for bilayer systems. These are $\pi-\pi$ transitions. Such transitions have been observed experimentally for doped $\pi$-conjugated polymers and supported by theoretical calculations for doped polythiophene.\cite{28,53} The calculated values of the $E_{binding}$ are found to be similar to those of the monolayer counterparts, which means that the bonding of the atoms is quite similar for monolayer and bilayer.

\begin{figure*}[htbp]
\includegraphics[width=0.9\textwidth,clip]{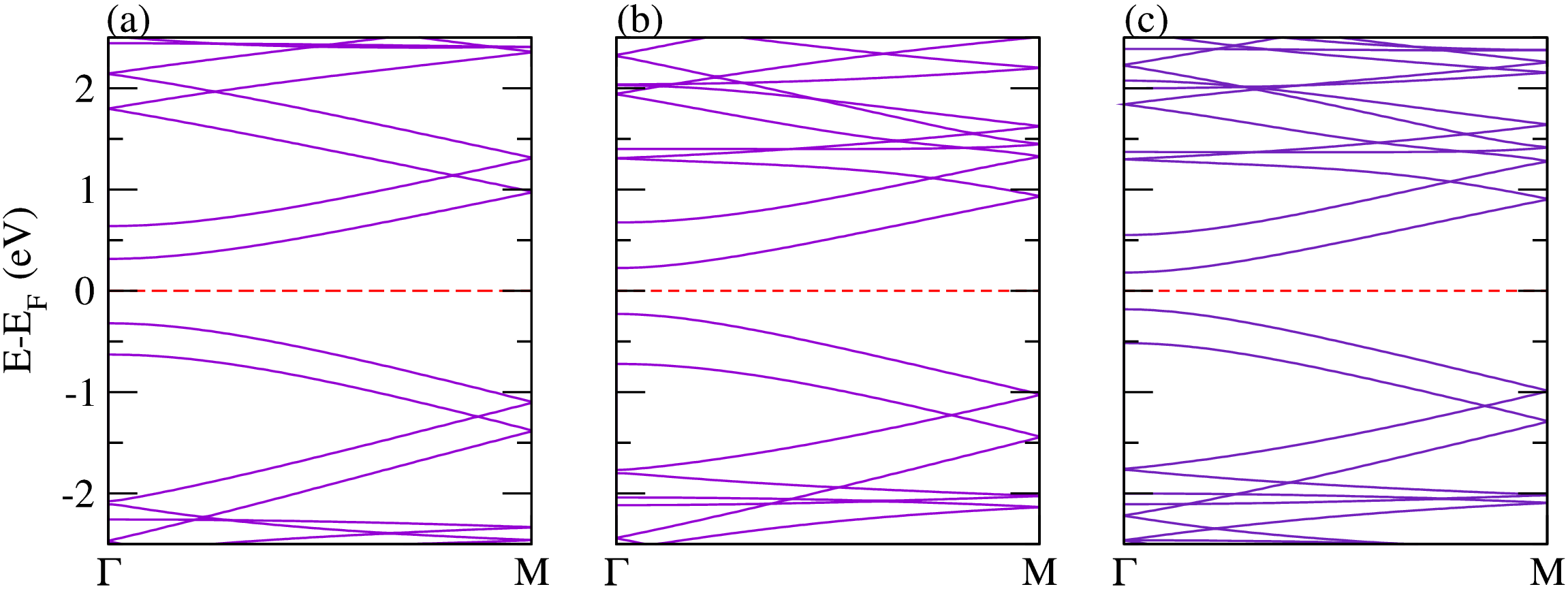}
\caption{The electronic structure of mixed bilayer systems: (a) polythiophene and polyselenophene, (b) polythiophene and polytellurophene, and (c) polyselenophene and polytellurophene.}
\end{figure*}

However, a significant variation in the valence and conduction band gap is found for mixed bilayers of polythiophene and polyselenophene or polythiophene and polytellurophene or polyselenophene and polytellurophene (compare Fig.\ 7(a-c)) because of the interaction between the two different layers. This can also be understood from the strong modification in the structural geometry (Fig.\ 4) as compared to the identical bilayers discussed above. On the other hand, the other parameters including bandwidths, band gap and $E_{binding}$ follow similar trends to those of the identical bilayers discussed above.

\begin{figure}[htbp]
\includegraphics[width=0.8\textwidth,clip]{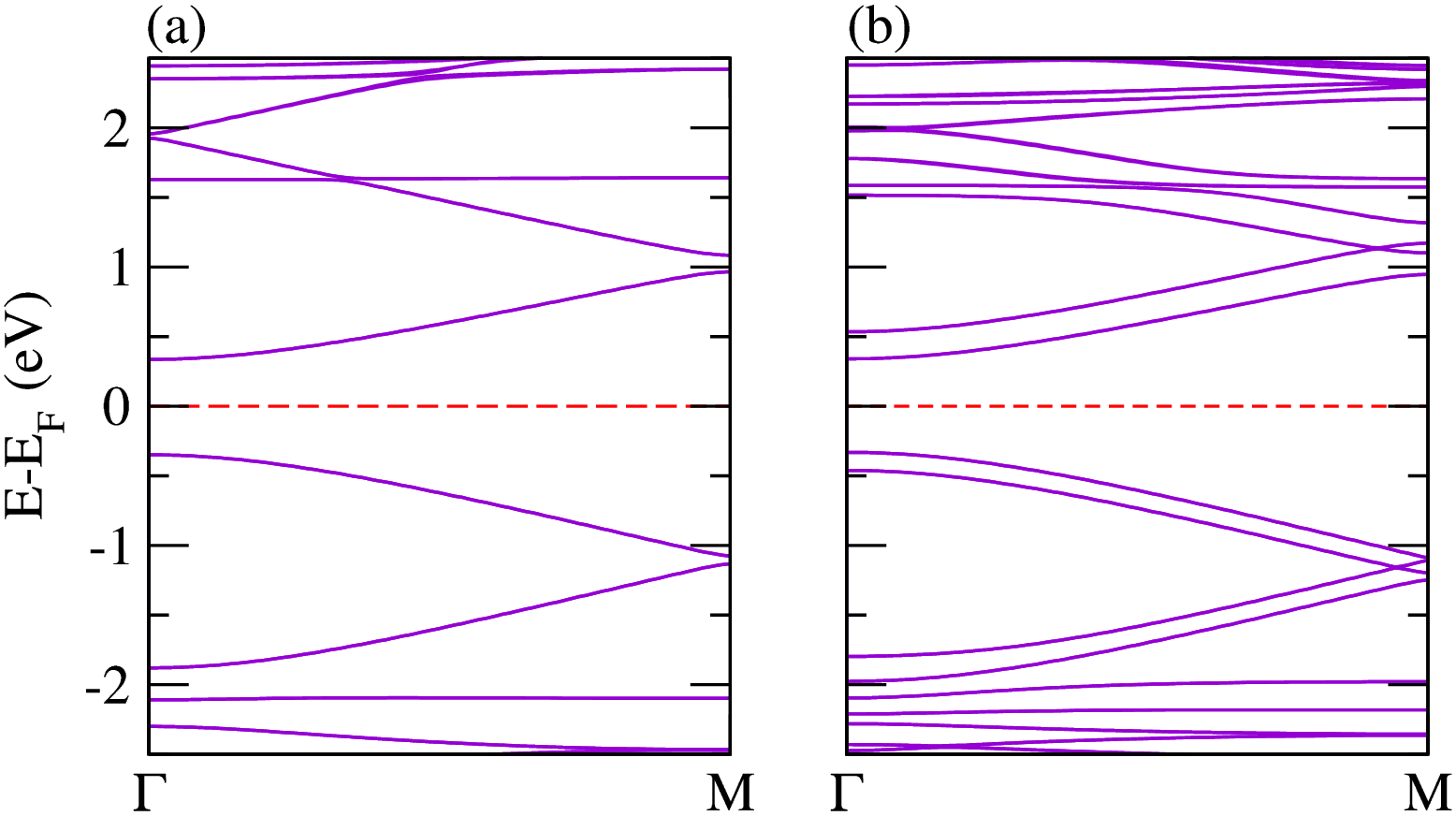}
\caption{The electronic structure of (a) S-Se-Te$^1$ (b) S-Se-Te$^2$.}
\end{figure}

In addition, we have investigated the electronic properties of the S-Se-Te$^1$ and S-Se-Te$^2$ systems, where the composition of these systems has been discussed above in the structural arrangements section. These systems have the highest values of the $E_{binding}$ among the systems under consideration. Hence, it is expected that such systems should be energetically very stable, and the constituent atoms should be bonded strongly in the pentagonal backbone. Thus, these systems could be utilized for device application, for example solar cells.\cite{102} All the calculated parameters are summarized in Tables 1 and 2. The electronic band structure is addressed in Fig.\ 8(a-b). The value obtained of the band gap is 0.70 eV (1.54 eV using B3LYP functional) for the monolayer and 0.64 eV (1.48 eV using B3LYP functional) for the bilayer with valence and conduction band gaps of 0.17 eV and 0.14 eV for S-Se-Te$^2$. Because of the interlayer interaction, we observe two bands immediately below and above the Fermi level, making the system promising for optical transitions.\cite{28}  

\subsection*{Molecular calculations}
The structures under consideration for the molecular calculations are addressed in Fig.\ 9 along with their corresponding HOMO and LUMO orbitals. Whereas, the HOMO and LUMO energies, HOMO-LUMO gap, $E_{binding}$, interlayer distance, buckling, bond distances, and bond angles for all the systems are summarized in Tables 3 and 4. Note that there are significant differences in the calculated values of the structural parameters, energies, and HOMO-LUMO gap between periodic and molecular calculation, however both approaches maintain internal consistency. These differences in the parameters can be understood due to the fact that in periodic calculations, the systems are fixed along the $y$- or $z$-directions, while in molecular calculations the structures are subjected to modifications in all three directions. Thus, in our opinion, molecular calculations provide slightly better results for comparing with the real experimental situation, especially regarding geometries, whereas, periodic structures are also important because electronic devices has been synthesized using crystalline structures.\cite{103} Note that, in case of mixed bilayers, the second layer is shifted as compared to the first layer, see Fig.\ 9. This indicates that in case of scheme (ii) (see section 3.1.1) a slipped ABA stacking pattern is energetically stable as compared to AAA non-slipped stacking pattern in case of scheme (i). In a recent experiment, both types of stacking patterns have been observed for two-dimensional $\pi$-conjugated polymers.\cite{104} The fact is that the unshifted layers are energetically less favourable, which means that this arrangement is only a local minimum on the potential energy surface for these particular systems.

\begin{figure}[htbp]
\includegraphics[width=0.8\textwidth,clip]{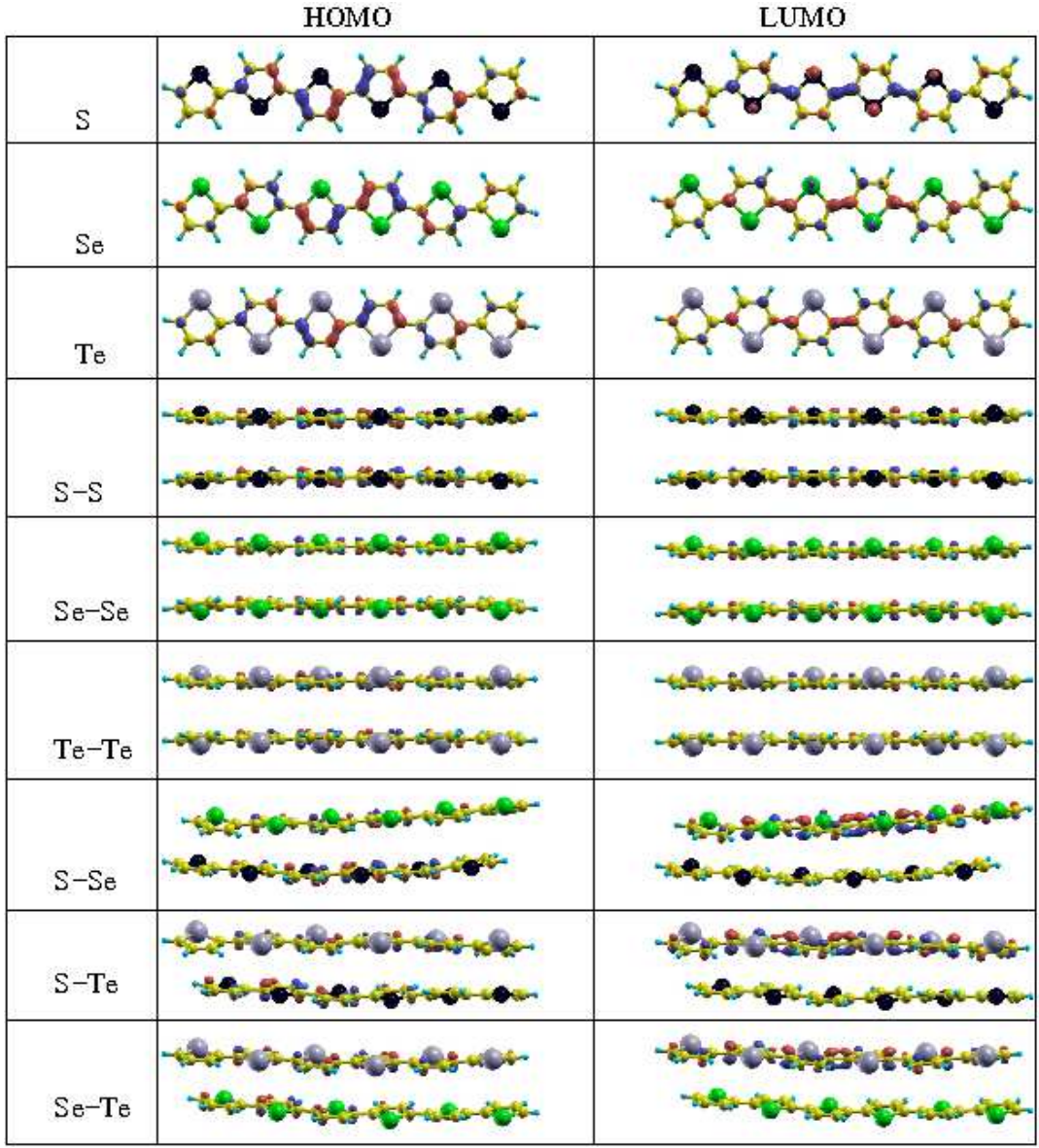}
\caption{The HOMO and LUMO of the systems under consideration, where yellow, blue, black, green, and gray balls represent C, H, S, Se, and Te atoms, respectively.}
\end{figure}

The calculated energy of the HOMO is slightly increasing (decreasing in magnitude) and that of LUMO decreasing (becoming more negative) from polythiophene to polytellurophene and a similar trend is achieved for the HOMO-LUMO gap, see Table 4. The values obtained of the HOMO-LUMO gap are higher (1.51 eV to 1.14 eV for polythiophene to polytellurophene) then those of the periodic approach due to the reasons as discussed above. The $E_{binding}$ is found to be significantly higher as compared to the periodic approach; it amounts to $-0.596$ eV, $-0.582$ eV, and $-0.569$ eV for polythiophene, polyselenophene, and polytellurophene, respectively. This clearly indicates that the constituent atoms are more strongly bonded with each other in a molecular form as compared to the crystalline form (periodic arrangement along x- and y-directions). The C$-$C bond lengths are again found to be shrinking upon substitution of the S atom in polythiophene by Se or Te. The C$-$S, C$-$Se, and C$-$Te bond lengths are found to be 1.75 \AA, 1.91 \AA, and 2.08 \AA, respectively. The bond length between C and the heteroatom increases with increasing atomic number due to the larger ionic size of the heteroatoms atoms, in good agreement with previous reports.\cite{88,105,106,107} 

The C$-$X$-$C bond angle is found decrease with increasing size of the heteroatoms (S, Se, and Te) and becomes 92$^\circ$ for polythiophene and 82$^\circ$ for polytellurophene. The other angles are found to be quite similar to those of the periodic approach, compare Table 1 and 3. Essentially, the obtained parameters are more or less following similar trends as their periodic counterparts. The values obtained of the band gap for polythiophene, polyselenophene, and polytellurophene agree well with previous experimental and theoretical reports.\cite{53,67,69,95} 

In case of bilayers composed of two identical layers of polythiophene, polyselenophene or polytellurophene, the averaged value of the interlayer separation amounts to 3.41 \AA\ to 3.53 \AA, see Table 3. We observed variable values of the interlayer separation because significant buckling is induced in both the layers due to the interlayer interactions. This agrees well with experimental observation in case of  $\pi$ stacked conjugated polymers.\cite{108,109} The buckling is found to be 0.25 \AA\ to 0.46 \AA\ for polythiophene to polytellurophene, respectively; such a buckling has already been observed for similar systems.\cite{53} The values obtained of the band gap are reduced significantly (1.16 eV for S, 1.07 eV for Se, and 0.89 eV for Te atoms in the backbone) as compared to those of their monolayer counterparts because of additional molecular orbitals that have developed close to the Fermi level. Similar behavior has been obtained previously for bilayer polythiophene with different chain lengths.\cite{53}

However, in the case of mixed bilayer (see Fig.\ 4), variable values of the interlayer spacing, buckling and bond lengths between C atoms and the heteroatoms are found. The reason for obtaining variable values is the interaction between the two different heteroatoms layers. The values obtained of the various parameters are listed in Table 3. The HOMO-LUMO gap is slightly higher then those of identical bilayers (Table 4). The variable value of the band gap which is close to 1.0 eV is promising for optical devices operated in low band gap regimes. Here, we obtain significant and predictable changes in the band gap of the polymers either by substituting heavier heteroatoms in the conjugated backbone or by creating bilayers. The significant lowering of the band gap (HOMO-LUMO gap) is due to the reduced aromaticity of a selenophene ring compared to a thiophene ring. The lower aromaticity in the main ring of the system lowers the HOMO-LUMO gap due to increased contributions of the quinoid structure of the main chain.\cite{67} A HOMO-LUMO gap lower then that of the monolayers and slightly higher then that of either of the bilayers is achieved for monolayer as well as bilayers of the systems where adjacent S atoms are substitutes by Se and Te atoms, the systems are also called S-Se-Te$^1$ and S-Se-Te$^2$, see Table 4. In summary, the calculations show that polythiophene-based systems can have wide ranges of the HOMO-LUMO gap (band gap), and they could be utilized for various electronic as well as optical devices accordingly. 
Therefore, synthesis of these systems would be useful.

\subsection*{Polypyrrole and its derivatives (Periodic approach)}
\subsubsection*{Structural arrangements}

The polypyrrole-based structures are presented in Fig.\ 3(b) and right panel of Fig.\ 4, where X refers to N, P, and As atoms. The supercell of the pristine monolayer of polypyrrole consists of 16 carbon (C) atoms, 12 Hydrogen (H) atoms, and 4 Nitrogen (N) or Phosphorus (P) or Arsenic (As) atoms. The calculated values of the structural parameters using periodic and molecular approaches are listed in Tables 1 and 3, respectively. For polypyrrole, the values obtained of the bond angles agree well with the available report.\cite{95} In case of polypyrrole, N atom is connected to each of the pentagonal rings of the backbone and one H atom, which is well explored.\cite{110} Experimentally, it has been suggested that the N atom can be replaced by a P or As atom.\cite{69} Therefore, we have extended our study by replacing N by P and As, such that these materials could be utilized in electronic devices such as optoelectronics.

For pristine polypyrrole, the buckling is found to be 0.11 \AA\ with a N$-$H bond length of 1.01 \AA. The C$-$C bond length is slightly longer than that of the polythiophene-based structures. The C$-$N bond length is 1.44 \AA, the C-X-C bond angle is 113$^{\circ}$, which is much larger than the C-X-C bond angle in the polythiophene counterpart, while the other bond angles are similar to those of polythiophene-based structures, see Table I. Moreover, buckling, X$-$H bond lengths, and C$-$X bond lengths are increasing when replacing N by P and As, see Table I. 

We also extended our study to two different types of bilayers systems, (i) both the layers are identical, 
i. e. N-N, P-P, and As-As, and (ii) the two layers are not identical i.e. N-P, N-As, and P-As, see Figure 2. Variable interlayer distances of 3.18-3.68 \AA, 1.73-2.69 \AA, and 1.73-2.72 \AA are obtained for the identical bilayer systems N-N, P-P, and As-As with a strong buckling, respectively. This indicates that there are strong structural modifications upon creating bilayers. In case of P and As monolayers or P-P and As-As bilayers, the P-H and As-H bonds point out-of-plane in order to get a stable structure, while the N-H bond points in-plane in case of N monolayer or bilayer of N-N. This is the reason why the interlayer distance in case of P-P and As-As bilayers is significantly smaller (1.73-2.72 \AA) than that of the N-N bilayer (3.18-3.68 \AA). Pristine polypyrrole bilayers (N-N) have been realized in experiment.\cite{111} 
At this point, we believe that the other bilayers should also be realized experimentally, which would widen the research area. Experimentally it has been proposed that the bilayers made of polpyrrole (N-N) are promising for applications in amperometric ion sensing.\cite{112} The buckling is found to be smaller and slightly less structural modifications are obtained in case of bilayers made of two non-identical layers (N-P, N-As, and P-As), see Table I and Fig.\ 4. Moreover, the electronic structure of the structures presented in right panel of Fig.\ 4 are addressed in Fig.\ 10.  Note that, in case of bilayers, the second layer is shifted as compared to the first layer, which indicates that the ABA kind of stacking is energetically favourable in these cases similar to bilayers made of polythiophene-based systems, as discussed above.\cite{104}

\subsubsection*{Electronic structure} 
\begin{figure}[htbp]
\includegraphics[width=1\textwidth,clip]{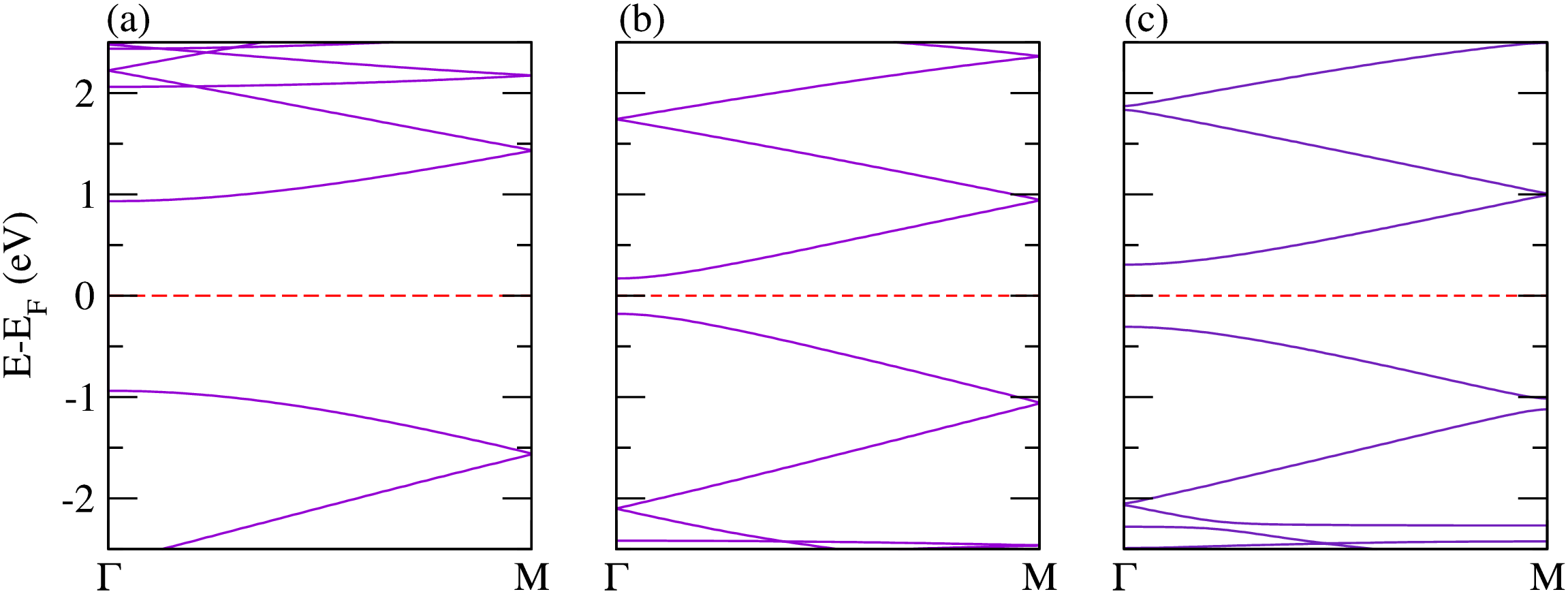}
\caption{The electronic structure of a single layer of pyrrole derivatives with (a) N, (b) P, and (c) As.}
\end{figure}

Polypyrrole is a wide band gap organic polymer with many applications, such as flexible batteries, semiconductors, large-area optical displays, and sensors.\cite{113} It has a larger band gap and hence, lower conductivity in the neutral state as compared to polythiophene. It has been reported that the effective mass of holes and electrons is comparatively higher than that of other polymers such as polythiophene, polyacetylene, and polyfuran.\cite{95} The calculated band structures of polypyrrole and its P and As analogues are presented in Figs.\ 10(a-c). The calculated values of the valence and conduction bandwidths along with the band gaps are summarized in Table II. The value obtained of the band gap is found to be 1.88 eV, which agrees well with a previously calculated value using first-principles calculations.\cite{95} The band gap shrinks significantly and becomes 0.34 eV and 0.61 eV for N atom replaced by P atom and As atom (see Figs.\ 8(b) and 8(c)), respectively. Moreover, the value of the band gap is found to be larger as compared to PBE functional by using the hybrid (B3LYP) functional, as expected, see Table II for the comparison. This fact is true for all the calculations for polypyrrole and its derivatives. As discussed above, such a replacement is quite feasible in a real experimental situation.\cite{71} A similar trend is found for the $E_{binding}$ as compared to the corresponding polythiophene-based monolayers.

\begin{figure}[htbp]
\includegraphics[width=1\textwidth,clip]{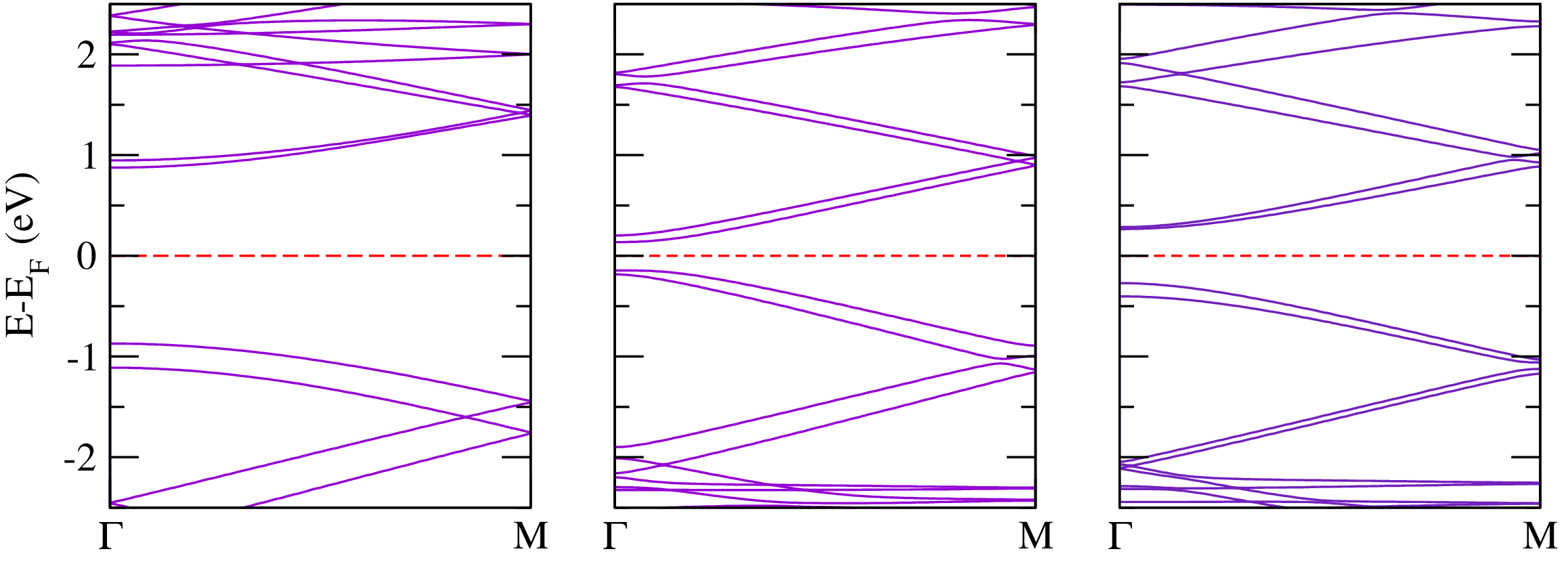}
\caption{The electronic structure of identical bilayers of (a) N-N, (b) P-P, and (c) As-As.}
\end{figure}

The calculated electronic band structures for bilayers, where both the layers are identical, are addressed in Fig.\ 11. It is clear from the figure that there are two bands immediately above and below the Fermi level with a finite gap, contributed from each of the layers. The valence and conduction bandwidths are growing from the N-N to As-As bilayers, while the band gap shrinks from N-N to As-As. A similar trend is also found to the corresponding polythiophene-based bilayers discussed above. The $E_{binding}$ again follows the same trend as for the corresponding monolayers. Because of strong structural deformation, non-identical bilayers made of N-P and P-As do not provide two bands immediately above the Fermi level; rather they provide two bands immediately below the Fermi level. Whereas, in case of the N-As bilayer, two bands immediately above and below the Fermi level are obtained, which is due to less modification in the structure, compare Fig.\ 12(a-c). Moreover, the other parameters including bandwidths, band gap and $E_{binding}$ have similar trends to those of the bilayers made of two identical layers.  

\begin{figure}[htbp]
\includegraphics[width=1\textwidth,clip]{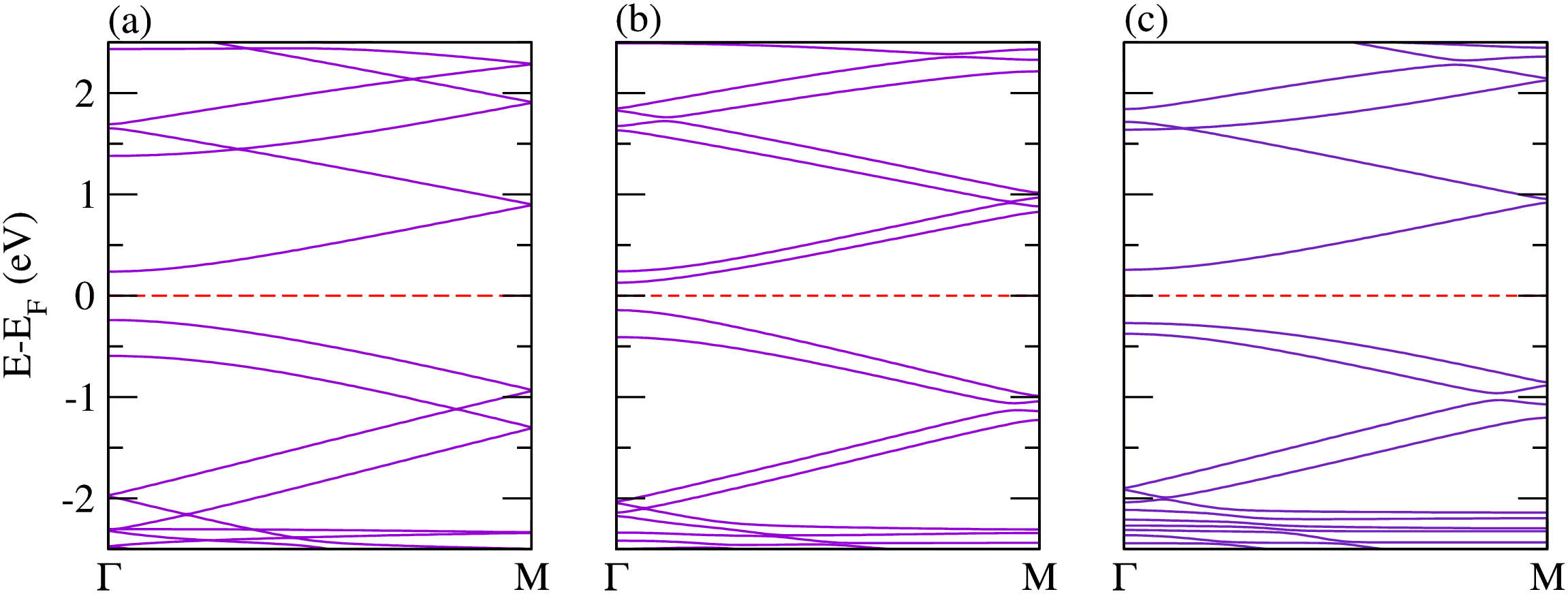}
\caption{The electronic structure of mixed polypyrrole derived bilayers, (a) N-P, (b) N-As, and (c) P-As.}
\end{figure}

\subsection*{Molecular calculations}

\begin{figure}[htbp]
\includegraphics[width=0.8\textwidth,clip]{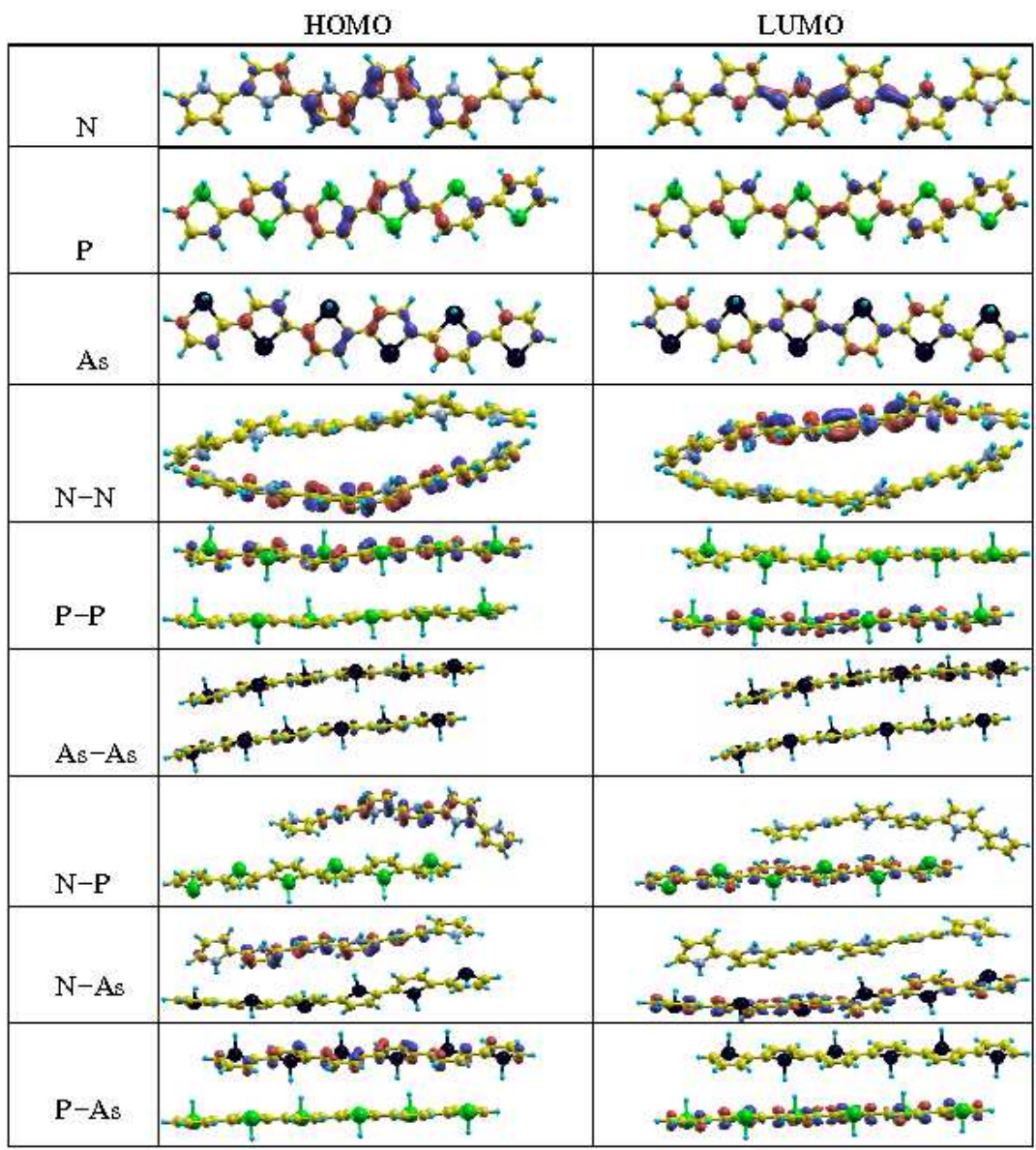}
\caption{The HOMO and LUMO of the systems under consideration.}
\end{figure}
The structures under consideration are shown in Fig.\ 13 with their corresponding HOMO and LUMO. The HOMO and LUMO energies, HOMO-LUMO gap, $E_{binding}$, distance between the two layers, buckling, bond distances, and bond angles for all the structures under investigation are presented in Table III. The electronic and structural parameters calculated using the molecular approach are slightly different from those of the periodic approach. The reason behind that is again the fact that in the periodic calculations, the systems are forced not to be moved along the $y$- or $z$-directions whereas in molecular calculations this is not the case. The values obtained of the HOMO and LUMO energies are significantly increasing in magnitude from N to P or As, which was not case for the polythiophene-based systems, compare Table IV. The value obtained of the HOMO-LUMO gap for polypyrrole (N) is found to be 2.25 eV and decreases to as low as 0.89 eV and 0.81 eV for N replaced by P and As in polypyrrole, respectively. The HOMO-LUMO gap is larger than that calculated using the periodic approach (1.88 eV) and close to the experimentally and theoretically obtained value of the HOMO-LUMO gap of 2.8 eV.\cite{95,114,115} 

The $E_{binding}$ is found to be significantly larger than that obtained from the periodic approach, indicating a strong bonding of atoms in this approach. The $E_{binding}$ are $-0.609$ eV, $-0.562$ eV, and $-0.548$ eV for N, P, and As systems, respectively, see Table II. The buckling is also irregular and higher than that of the periodic approach. Whereas, other structural parameters such as bond lengths and bond angles are found to be similar to the values obtained using the periodic approach. For bilayers (made of the two identical layers), a slight reduction in the HOMO and LUMO energies is achieved with slightly smaller HOMO-LUMO gaps of 2.20 eV, 0.89 eV, and 0.76 eV for bilayers of N-N, P-P, and As-As, respectively. Note that the HOMO is contributed by one of the layers and the LUMO is contributed by the other layer in case of N-N and P-P, while in case of As-As, the HOMO and LUMO both are contributed from both the layers, in agreement with previous findings.\cite{53} Due to the very strong structural modification and corrugation, we observed variable values of the interlayer separation with significant buckling; this agrees well with experimental observation in case of $\pi$ stacked conjugated polymers as discussed above.\cite{108,109} The trends for the other structural parameters remain similar to those obtained from periodic calculations.

Moreover, in case of mixed bilayers N-P, N-As, and P-As (Fig.\ 10), the N-P bilayer system has significantly larger LUMO energy as compared to N-N but N-As and P-As bilayers have similar trends of HOMO or LUMO to the P-P and As-As bilayers. The value obtained of the HOMO-LUMO gap is strongly reduced (0.41 eV, 0.40 eV , and 0.73 eV, for N-P, N-As, and P-As bilayers, respectively) as compared to their identical bilayer counterparts (see Table II). The strong reduction in the HOMO-LUMO gap can be understood from the different interlayer interactions and structural modifications. Again in these bilayers, the HOMO is contributed from one layer and the LUMO is contributed from the other layer. The $E_{binding}$ is slightly smaller than that of the identical counterparts, and larger interlayer separation and smaller buckling are found. Variable X$-$H bonds are found due to strong structural modification. All other structural parameters are found to be following similar trend as compared to N-N, P-P, and As-As bilayers.

\section*{Bilayer of polythiophene and polypyrrole}
In addition, we further studied the structural and electronic properties of mixed bilayers made of polythiophene and polypyrrole (S-P). The structural and electronic parameters using both the periodic and molecular approaches are summarized in Tables I and II. Valence and conduction bandwidths of 0.60 eV and 0.67 eV, a band gap of 0.72 eV using PBE functional and 1.72 eV using hybrid (B3LYP) functional, a band gap immediately above the Fermi level of 1.03 eV, and a band gap immediately below the Fermi level of 0.41 eV are obtained. Other structural parameters are similar to those of other systems under study. A HOMO-LUMO gap of 0.87 eV is found in the molecular calculation, which is close to that obtained in the periodic calculation (0.72 eV), see Tables I and II. Other structural parameters (bond lengths and bond angles) have again similar trends as other systems under study. It should be noted that such systems are promising to achieve low band gap $\pi$-conducting polymers, and could therefore be utilized in electronic devices.\cite{91,92}

\section*{Conclusion}
Using first-principles calculations, the structural and electronic properties (the band gap modulation) of polythiophene and polyprrrole-based systems (in periodic as well as in oligomer approaches) have been investigated. The effects in the structural and electronic properties for S substituted by Se and Te in case of polythiophene and N substituted by P and As for polypyrrole have been addressed. Furthermore, the effect of the stacking in pure polythiophene as well as in pure polypyrrole have been studied and a linear reduction of the band gap or HOMO-LUMO gap as function of the stacking was found. In addition, we also employed hybrid functional based calculations for periodic cases in order to understand the effect of the hybrid functional on the band gap. It was found that the band gap is larger using hybrid (B3LYP) calculations as expected, while the trend in the band gap is similar to that of the PBE calculations.

This study has also been extended to polythiophene and polytellurophene and to polypyrrole derivatives where the N atom is substituted by P and As atoms. In addition, we performed $\Gamma$-point phonon calculation for selected systems to check the stability of these systems. The values obtained of the $E_{binding}$ suggest that all the systems under consideration are thermodynamically stable and can therefore be realized experimentally. The trends in the $E_{binding}$ as a function of stacking as well as substitution remain similar for both the periodic and molecular approaches. Moreover, the bilayer constructed from polythiophene on top of polypyrrole has also been investigated. It is believed that the comprehensive analysis of such investigation will be useful for fundamental understanding in the structural and electronic properties and the band gap or HOMO-LUMO gap, which could be very important way to construct polythiophene and polypyrrole-based electronic devices because of wide range of the band gap can be obtained.

%\section*{References}

\pagebreak
\textbf{Acknowledgement}\\
\noindent G. S. acknowledges funding from the Natural Sciences and Engineering Council of Canada (NSERC, Discovery Grant). M. S. F. acknowledges support by the Natural Sciences and Engineering Research Council (NSERC) of Canada, the Canada Research Chair program, Canada Foundation for Innovation (CFI), the Manitoba Research and Innovation Fund, and the University of Manitoba.\\

\noindent \textbf{Contributions}\\
\noindent T.P.K. performed the calculations. T.P.K., G.S. and M.S.F wrote the manuscript.\\

\noindent \textbf{Competing interests}\\
\noindent The authors declare no competing financial interests.\\

\begin{table}[h]
\begin{tabular}{|c|c|c|c|c|c|c|}
\hline
System&X$-$H&C$-$C &C$-$X &C-X-C&C-C-C &X-C-C \\
\hline
 \multicolumn{7}{|c|}{\multirow{1}{*}{\textbf{Polythiophene derivatives, monolayers}}}\\
\cline{1-7}
\hline
S&--&1.38-1.41&1.73&93&113-128&109-119   \\
\hline
Se&--&1.37-1.40&1.88&85&114-129 &112-117    \\
\hline
Te&--&1.35-1.38&2.08&76&115-127&116-119     \\
\hline
 \multicolumn{7}{|c|}{\multirow{1}{*}{\textbf{Bilayer (identical layers) polythiophene with S, Se, and Te}}}\\
\cline{2-7}
\hline
S-S&--&1.38-1.43&1.74&92&113-129&109-121 \\
\hline
Se-Se&--&1.39-1.44&1.88&85&113-128&112-117  \\
\hline
Te-Te&--&1.37-1.45&2.07&76& 116-127&116-119      \\
\hline
 \multicolumn{7}{|c|}{\multirow{1}{*}{\textbf{Bilayer (non-identical layers) polythiophene with S, Se, and Te}}}\\
\cline{2-7}
\hline
S-Se&--&1.37-1.39&1.73-1.88&85-92&114-129&110-121 \\
\hline
S-Te&--&1.36-1.38&1.74-2.06&76-92&113-128&111-120 \\
\hline
Se-Te&--&1.36-1.38&1.88-2.07&76-85&114-128&112-116 \\
\hline
 \multicolumn{7}{|c|}{\multirow{1}{*}{\textbf{Polythiophene with alternative S is replaced by Se and Te}}}\\
\cline{2-7}
\hline
S-Se-Te$^1$&--&1.38-1.40&1.72-2.08&79-91&117-129&112-120\\
\hline
S-Se-Te$^2$&--&1.39-1.41&1.71-2.07&79-92&115-128&111-119 \\
\hline
 \multicolumn{7}{|c|}{\multirow{1}{*}{\textbf{Monolayer polypyrrole with N, P, and As}}}\\
\cline{2-7}
\hline
N&1.01&1.40-1.55&1.44&113&110-128&102-128   \\
\hline
P&1.43&1.37-1.42&1.82&88&114-126&110-122    \\
\hline
As&1.55&1.36-1.42&1.99&85&117-126&110-123    \\
\hline
 \multicolumn{7}{|c|}{\multirow{1}{*}{\textbf{Bilayer (identical layers) polypyrrole with N, P, and As}}}\\
\cline{2-7}
\hline
N-N&1.01&1.42-1.55&1.45&112&110-128&102-129       \\
\hline
P-P&1.54&1.37-1.42&1.99&84&117-126&109-129       \\
\hline
As-As&1.54&1.38-1.42&1.99&84&117-125&109-123       \\
\hline
 \multicolumn{7}{|c|}{\multirow{1}{*}{\textbf{Bilayer (non-identical layers) polypyrrole with N, P, and As}}}\\
\cline{2-7}
\hline
N-P&1.01-1.43&1.40-1.55&1.45-1.82&89-113&110-128&102-128\\
\hline
N-As&1.02-1.55&1.36-1.42&1.45-1.48&84-113&110-128&102-128 \\
\hline
P-As&1.43-1.54&1.37-1.42&1.83-1.99&84-89&114-126&109-123 \\
\hline
 \multicolumn{7}{|c|}{\multirow{1}{*}{\textbf{Bilayer composed of pure polythiophene and polypyrrole}}}\\
\cline{2-7}
\hline
S-P&1.02&1.39-1.42&1.44-1.74&93-113&110-129&110-128    \\
\hline
\end{tabular}
\caption{Structural parameters, periodic calculations: selected bond lengths (in \AA) and bond angles (in $^\circ$).} 
\end{table}

\begin{table}[ht]
\begin{tabular}{|c|c|c|c|c|c|c|c|c|c|}
\hline
System&V$_{BW}$ &C$_{BW}$ &$E_{gap}$ (PBE)&$E_{gap}$ (B3LYP) &$E_{gap}^{VB}$ &$E_{gap}^{CB}$ &$E_{binding}$ &$d_{int}$ &Buckling  \\
\hline
 \multicolumn{9}{|c|}{\multirow{1}{*}{\textbf{Polythiophene derivatives, monolayers}}}\\
\cline{1-9}
\hline
S&0.74&0.63&1.12&2.08&-- &--&$-0.483$&--&0.10   \\
\hline
Se&0.76&0.69&0.68&1.53&--&--&$-0.462$&--&0.12   \\
\hline
Te&0.74&0.81&0.26&0.97&--&--&$-0.455$&--&0.17    \\
\hline
 \multicolumn{9}{|c|}{\multirow{1}{*}{\textbf{Bilayer (identical layers) polythiophene with S, Se, and Te}}}\\
\cline{2-9}
\hline
S-S&1.14&0.62&0.83&1.81&0.14&0.21&$-0484$&3.33-3.43&0.11 \\
\hline
Se-Se&0.75&0.70&0.51&1.33&0.16&0.24&$-0.463$&3.36-3.44&0.12  \\
\hline
Te-Te&0.79&0.73&0.22&0.81&0.31&0.40&$-0.457$&3.36-3.42&0.16 \\
\hline
 \multicolumn{9}{|l|}{\multirow{1}{*}{\textbf{Bilayer (non-identical layers) polythiophene with S, Se, and Te}}}\\
\cline{2-9}
\hline
S-Se&0.77&0.63&0.64&1.51&0.31&0.29&$-0.476$&3.42-3.62&0.46 \\
\hline
S-Te&0.75&0.73&0.44&1.29&0.49&0.44&$-0.469$&3.39-3.65&0.72 \\
\hline
Se-Te&0.78&0.72&0.36&1.19&0.32&0.40&$-0.462$&3.39-3.68&0.84 \\
\hline
 \multicolumn{9}{|c|}{\multirow{1}{*}{\textbf{Polythiophene with alternative S replaced by Se and Te}}}\\
\cline{2-9}
\hline
S-Se-Te$^1$&0.61&0.63&0.70&1.54&--&--&$-0.498$&--&0.12\\
\hline
S-Se-Te$^2$&0.77&0.65&0.64&1.48&0.17&0.14&$-0.496$&3.48-3.75&0.74 \\
\hline
 \multicolumn{9}{|c|}{\multirow{1}{*}{\textbf{Monolayer polypyrrole with N, P, and As}}}\\
\cline{2-9}
\hline
N&0.61&0.47&1.88&3.01&--&--&$-0.492$&--&0.11   \\
\hline
P &0.88&0.79&0.34&1.27&--&--&$-0.461$&--&0.17  \\
\hline
As&0.67&0.72&0.61&1.58&--&--&$-0.448$&--&1.91    \\
\hline
 \multicolumn{9}{|c|}{\multirow{1}{*}{\textbf{Bilayer (identical layers) polypyrrole with N, P, and As}}}\\
\cline{2-9}
\hline
N-N&0.71&0.61&0.86&2.58&0.13&0.26&$-0.494$&3.18-3.68&1.52 \\
\hline
P-P&0.74&0.69&0.51&1.49&0.15&0.22&$-0.464$&1.73-2.69&1.87       \\
\hline
As-As&0.77&0.73&0.22&1.08&0.30&0.40&$-0.450$&1.73-2.72&1.89 \\
\hline
 \multicolumn{9}{|c|}{\multirow{1}{*}{\textbf{Bilayer (non-identical layers) polypyrrole with N, P, and As}}}\\
\cline{2-9}
\hline
N-P&0.27&0.70&0.46&1.81&0.35&--&$-0.491$&2.31-2.67&0.81\\
\hline
N-As&0.81&0.70&0.29&1.17&0.24&0.12&$-0.462$&1.51-3.98&0.99-2.24 \\
\hline
P-As&0.57&0.64&0.54&1.88&0.08&--&$-0.453$&1.74-2.01&0.51-1.64\\
\hline
 \multicolumn{9}{|c|}{\multirow{1}{*}{\textbf{Bilayers composed of pure polythiophene and polypyrrole}}}\\
\cline{2-9}
\hline
S-P&0.60&0.67&0.72&1.72&0.41&1.03&$-0.491$&2.29-3.76&1.69  \\
\hline
\end{tabular}
\caption{Periodic calculations: the calculated values of the valence and conduction bandwidths (in eV), band gap (in eV), band gap in valence and conduction bands at the $\Gamma$ point (in eV), binding energy (in eV), interlayer spacing (in \AA), and buckling (in \AA), where X = S, Se, Te, N, P, and As and S-Se-Te$^1$ and S-Se-Te$^2$ correspond to single layer and bilayers.} 
\end{table}

\begin{table}[h]
\begin{tabular}{|c|c|c|c|c|c|c|c|c|c|}
\hline
System&X$-$H&C$-$C&C$-$X&C-X-C&C-C-C&X-C-C \\
\hline
 \multicolumn{7}{|c|}{\multirow{1}{*}{\textbf{Polythiophene derivatives, monolayers}}} \\
\cline{1-7}
\hline
S&--&1.40-1.44&1.75& 92 &113-129&110-120   \\
\hline
Se&--&1.40-1.42&1.91&87&115-129&111-120   \\
\hline
Te&--&1.38-1.41&2.08&82&118-128&108-122 \\
\hline
 \multicolumn{7}{|c|}{\multirow{1}{*}{\textbf{Bilayer (identical layers) polythiophene with S, Se, and Te}}} \\
\cline{2-7}
\hline
S-S&--&1.39-1.44&1.75&91&114-128&111-120\\
\hline
Se-Se&--&1.38-1.43&1.90&87&115-129&109-120  \\
\hline
Te-Te&--&1.38-1.42&2.12&82&118-128&109-122      \\
\hline
 \multicolumn{7}{|c|}{\multirow{1}{*}{\textbf{Bilayer (non-identical layers) polythiophene with S, Se, and Te}}}\\
\cline{2-7}
\hline
S-Se&--&1.38-1.42&1.73-1.91&87-91&113-129&110-121 \\
\hline
S-Te&--&1.38-1.43&1.75-2.12&81-92&113-128&111-121 \\
\hline
Se-Te&--&1.38-1.43&1.91-2.12&81-87&116-129&109-122\\
\hline
 \multicolumn{7}{|c|}{\multirow{1}{*}{\textbf{Polythiophene with alternative S is replaced by Se and Te}}}\\
\cline{2-7}
\hline
S-Se-Te$^1$&--&1.41-1.43&1.73-1.12&82-83&114-128&109-122\\
\hline
S-Se-Te$^2$&--&1.40-1.42&1.74-2.11&81-84&113-127&111-123\\
\hline
 \multicolumn{7}{|c|}{\multirow{1}{*}{\textbf{Monolayer polypyrrole with N, P, and As}}}\\
\cline{2-7}
\hline
N&1.01&1.42-1.44&1.38&111&108-132&107-122 \\
\hline
P &1.44&1.36-1.43&1.85&90&114-127&110-124  \\
\hline
As&1.54&1.37-1.43&1.99&86&117-127&109-124 \\
\hline
 \multicolumn{7}{|c|}{\multirow{1}{*}{\textbf{Bilayer (identical layers) polypyrrole with N, P, and As}}}\\
\cline{2-7}
\hline
N-N&1.01&1.39-1.44&1.38&110&108-131&106-122 \\
\hline
P-P&1.43&1.38-1.42&1.85&90&115-127&108-124 \\
\hline
As-As&1.54&1.39-1.41&1.99&86&117-127&108-124\\
\hline
 \multicolumn{7}{|c|}{\multirow{1}{*}{\textbf{Bilayer (non-identical layers) polypyrrole with N, P, and As}}}\\
\cline{2-7}
\hline
N-P&1.02-1.73&1.39-1.44&1.38-1.85&91-110&115-127&106-124\\
\hline
N-As&1.01-1.54&1.39-1.44&1.96-1.38&85-110&108-131&107-1.21\\
\hline
P-As&1.43-1.54&1.39-1.42&1.85-1.99&86-90&114-126&108-123\\
\hline
 \multicolumn{7}{|c|}{\multirow{1}{*}{\textbf{Bilayer composed of pure polythiophene and polypyrrole}}}\\
\cline{2-7}
\hline
S-P&1.01&1.37-1.43&1.38-1.73&92-110&107-131&107-121     \\
\hline
\end{tabular}
\caption{Structural parameters, molecular calculations: selected bond lengths (in \AA) and bond angles (in $^\circ$).} 
\end{table}

\begin{table}[h]
\begin{tabular}{|c|c|c|c|c|c|c|c|c|c|c|c|c|c|}
\hline
System&HOMO&LUMO &HOMO-LUMO &$E_{binding}$&$d_{int}$ &Buckling\\
\hline
 \multicolumn{7}{|c|}{\multirow{1}{*}{\textbf{Polythiophene derivatives, monolayers}}}\\
\cline{2-7}
\hline
S&$-4.547$&$-3.034$&1.51&$-0.596$&--&0.00 \\
\hline
Se&$-4.535$&$-3.226$&1.31&$-0.582$&--&0.00 \\
\hline
Te&$-4.516$&$-3.377$&1.14&$-0.569$&--&0.00\\
\hline
 \multicolumn{7}{|c|}{\multirow{1}{*}{\textbf{Bilaye Bilayer (identical layers) polythiophene with S, Se, and Te}}}\\
\cline{2-7}
\hline
S-S&$-4.273$&$-3.112$&1.16&$-0.601$&3.41-3.53&0.25\\
\hline
Se-Se&$-4.335$&$-3.261$&1.07&$-0.587$&3.44-3.61&0.35\\
\hline
Te-Te&$-4.277$&$-3.381$&0.89&$-0.575$&3.39-3.52&0.46\\
\hline
 \multicolumn{7}{|c|}{\multirow{1}{*}{\textbf{Bilayer Bilayer (non-identical layers) polythiophene with S, Se, and Te}}}\\
\cline{2-7}
\hline
S-Se&$-4.308$&$-3.118$&1.19&$-0.594$&3.33-3.49&0.15-0.89\\
\hline
S-Te&$-4.380$&$-3.279$&1.10&$-0.588$&3.31-3.50&0.40-0.79\\
\hline
Se-Te&$-4.345$&$-3.303$&1.04&$-0.577$&3.41-3.58&0.42-0.65\\
\hline
 \multicolumn{7}{|c|}{\multirow{1}{*}{\textbf{Polythiophene with alternative S is replaced by Se and Te}}}\\
\cline{2-7}
\hline
S-Se-Te$^1$&$-4.533$&$-3.203$&1.23&$-0.582$&--&0.00\\
\hline
S-Se-Te$^2$&$-4.511$&$-3.136$&1.37&$-0.579$&3.42-3.51&0.23-0.27\\
\hline
 \multicolumn{7}{|c|}{\multirow{1}{*}{\textbf{Monolayer polypyrrole with N, P, and As}}}\\
\cline{2-7}
\hline
N&$-3.936$&$-1.687$&2.25&$-0.609$&--&0.09-0.12 \\
\hline
P &$-4.464$&$-3.570$&0.89&$-0.562$&--&0.25-1.12 \\
\hline
As&$-4.429$&$-3.617$&0.81&$-0.548$&--&0.27-1.14\\
\hline
 \multicolumn{7}{|c|}{\multirow{1}{*}{\textbf{Bilayer (identical layers) polypyrrole with N, P, and As}}}\\
\cline{2-7}
\hline
N-N&$-3.945$&$-1.745$&2.20&$-0.610$&0.96-5.53&1.72-3.93 \\
\hline
P-P&$-4.454$&$-3.591$&0.86&$-0.599$&0.75-3.57&0.60-0.71 \\
\hline
As-As&$-4.366$&$-3.605$&0.76&$-0.588$&3.02-3.44&0.41-0.48\\
\hline
 \multicolumn{7}{|c|}{\multirow{1}{*}{\textbf{Bilayer (non-identical layers) polypyrrole with N, P, and As}}}\\
\cline{2-7}
\hline
N-P&$-3.995$&$-3.580$&0.41&$-0.586$&2.17-2.73&0.87-3.38\\
\hline
N-As&$-3.992$&$-3.592$&0.40&$-0.582$&1.45-3.36&0.38-2.32\\
\hline
P-As&$-4.410$&$-3.682$&0.73&$-0.578$&2.43-2.89&0.66-3.21\\
\hline
 \multicolumn{7}{|c|}{\multirow{1}{*}{\textbf{Bilayer composed of pure polythiophene and polypyrrole}}}\\
\cline{2-7}
\hline
S-P&$-3.931$&$-3.057$&0.87&$-0.589$&3.55-3.75&0.76-0.98   \\
\hline
\end{tabular}
\caption{Molecular calculations: The calculated values of the highest occupied (HO) and lowest unoccupied (LUMO) molecular orbitals (in eV), HOMO-LUMO (in eV), binding energy (in eV), interlayer spacing (in \AA), buckling (in \AA), bond lengths (in \AA), and bond angles (in $^\circ$).} 
\end{table}

\end{document}